\documentstyle[aps,prl,floats]{revtex}
\begin{document}
\draft
\twocolumn[\hsize\textwidth\columnwidth\hsize\csname @twocolumnfalse\endcsname

\title
{\bf How to create Alice string (half-quantum vortex) in a vector Bose-Einstein
condensate.}
\author{ U. Leonhardt$^{1,2}$
\& G.E. Volovik$^{3,4}$ }
\address{$^1$ School of Physics and Astronomy, University of St. Andrews,
North Haugh, St. Andrews, Fife, KY16 9SS, Scotland\\
$^2$ Department of Physics, Royal Institute of Technology,
Lindtstedtsv\"agen 24, 100 44 Stockholm, Sweden\\
$^3$ Low Temperature Laboratory, Helsinki University of
Technology, 02015 Espoo, Finland\\
$^4$ Landau Institute for Theoretical Physics, 117334
Moscow, Russia. }

\date{\today} \maketitle
\begin{abstract}{
We suggest a procedure how to prepare the vortex with $N=1/2$ winding number
-- the counterpart of the Alice string -- in  Bose--Einstein
condensates. }
\end{abstract}
\

\pacs{PACS numbers:     }

\
]

Vortices with fractional winding number can exist in different condensed
matter systems, see review paper \cite{Monopoles}. Observation of
atomic Bose-condensates with multi-component order parameter in laser
manipulated
traps opens the possibility to create half-quantum vortices there. We discuss
the $N=1/2$ vortices in the Bose-condensate with the
hyperfine spin $F=1$, and also in the mixture of two Bose-condensates.

The order parameter of
$F=1$ Bose-condensate consists of 3 complex components according to the number
of the projections
$M=(+1,0,-1)$. These components can be organized to form the complex vector
${\bf a}$:
\begin{equation}
\Psi_\nu=\left(
\matrix{
\Psi_{+1}\cr
\Psi_{0}\cr
\Psi_{-1}\cr
}\right)= \left(
\matrix{
{a_x+ia_y\over \sqrt{2}}\cr
a_z\cr
{a_x-ia_y\over \sqrt{2}}\cr
}\right)
\,\,.
\label{ComponentRepresentation}
\end{equation}
 There are two symmetrically distinct phases
of the $F=1$ Bose-condensates:

(i) The chiral or ferromagnetic state occurs when the scattering length
$a_2$ in
the
scattering channel of two atoms with the total spin 2 is less than that with
the total spin zero, $a_2<a_0$ \cite{Ho,TheoryVortexFormation}.
It is described by the complex vector
\begin{equation}
{\bf a}=f(\hat{\bf m} + i \hat{\bf n})~,
\label{Complexa}
\end{equation}
 where $\hat{\bf m}$ and
$\hat{\bf n}$ are mutually orthogonal unit vector with $\hat{\bf l}=\hat{\bf
m}\times \hat{\bf n}$ being the direction of the spontaneous momentum ${\bf
F}$ of
the Bose condensate, which violates the parity and time reversal symmetry;
$f$ is
the amplitude of the order parameter.

(ii) The polar or superfluid nematic state,
which occurs  for $a_2>a_0$, is
described by the real vector up to the phase factor
\begin{equation}
{\bf a}=f\hat{\bf d}e^{i\Phi}~,
\label{Reala}
\end{equation}
where $\hat{\bf d}$ is a real unit vector. The direction of the vector
$\hat{\bf d}$ can be inverted by the change of the phase $\Phi \rightarrow \Phi
+\pi$. That is why phase-insensitive properties of the polar state are also
insensitive to the reversal of the direction of $\hat{\bf d}$. In this respect
$\hat{\bf d}$ is similar to the director in nematic liquid crystals.

The chiral state (i) corresponds to the orbital part of the matrix order
parameter in superfluid $^3$He-A, while the nematic state (ii) corresponds
to the
spin part of the same $^3$He-A order parameter. The order parameter matrix
of $^3$He-A is the product of two vector order parameters: $A_{\alpha k}
\propto
a_\alpha^{\rm nematic}a_k^{\rm chiral}$. That is why each of the two states
shares some definite properties of superfluid
$^3$He-A.

In particular the chiral state (i) displays continuous vorticity
\cite{Ho,TheoryVortexFormation}, which was heavily investigated in superfluid
$^3$He-A (see \cite{Blaauwgeers} and reviews
\cite{EltsovKrusius,SalomaaVolovik1987}). An isolated continuous vortex is the
so called Anderson-Toulouse-Chechetkin vortex.  The smooth core of the vortex
represents the skyrmion, in which the $\hat{\bf l}$-vector sweeps the whole
unit
sphere. Outside the soft core the $\hat{\bf l}$-vector is
uniform, while the order parameter phase  has finite winding. In $^3$He-A and
thus in the $F=1$ Bose-condensate too, it is the
$4\pi$ winding around the soft core, i.e. the continuous vortex  has
winding number $N=2$. This continuous  vortex can be also represented
\cite{SalomaaVolovik1987,EltsovKrusius}  as a pair of the so called continuous
Mermin-Ho vortices \cite{Mermin-Ho}, each having the winding number
$N=1$. The $\hat{\bf l}$-vector in the Mermin-Ho vortex covers only half of a
unit sphere and thus is not uniform outside the soft core. Such a half-skyrmion
is also called the meron. An optical method to create the meron -- the
Mermin-Ho
vortex -- in the $F=1$ Bose-condensate has been recently discussed in
Ref. \cite{Marzlin}.

For the spin-$1/2$ Bose-condensates the order parameter is a spinor, which
represents the ``half of the vector''. That is why the continuous
Anderson-Toulouse-Chechetkin vortex (in which the $\hat{\bf l}$ sweeps the
whole
unit sphere) has twice less winding number in such condensates,  i.e. the
skyrmion is the $N=1$ continuous vortex \cite{HoTalk}. The spinorial order
parameter is the counterpart of the order parameter in the Standard Model
of the
electroweak interactions, which is the spinor Higgs field transforming under
$SU(2)$ symmetry group. That is why the $N=1$ continuous vortex in the
spin-$1/2$
Bose-condensates simulates the continuous electroweak string in the Standard
Model. The  Higgs field in the continuous electroweak string (and thus the
$N=1$
continuous vortex in the spin-$1/2$ Bose-condensate) has the following
distribution of the order parameter
\cite{HindmarshKibble,AchucarroVachaspati}:
\begin{equation}
\left(
\matrix{
\Psi_\uparrow\cr
\Psi_\downarrow\cr
}\right)=f\left(
\matrix{ e^{i\phi} \cos {\theta(r)\over 2}
\cr
\sin {\theta(r)\over 2}\cr
}\right)
\, ~~, ~~\hat{\bf l}=\hat{\bf
z} \cos \theta(r)+\hat{\bf
r} \sin \theta(r) \,.
\label{Skyrmion}
\end{equation}
Here $(z,r,\phi)$ are coordinates of cylindrical system; $\theta(0)=\pi$; and
$\theta(\infty)=0$. Note that the meron configuration in such system,  with
$\theta(0)=\pi$ and $\theta(\infty)=\pi/2$,
would have
$N=1/2$ winding number.

The $N=1$ vortices with the order parameter described by Eq.(\ref{Skyrmion}),
have been recently generated in the Bose-condensate with two internal levels
\cite{Matthews}, following the proposal elaborated in Ref.\cite{Williams}.
Though these two internal levels are not related by an exact $SU(2)$ symmetry,
under some conditions there is an approximate $SU(2)$ symmetry, and the
$N=1$ vortex does represent a skyrmion. This vortex has a smooth (soft) core,
which size is essentially larger than that of the conventional
vortex core which has the dimension of order of the coherence length. Such
enhancement of the core size allowed for the observation of the $N=1$
vortex-skyrmion by optical methods
\cite{Matthews}. From the Eq.(\ref{Skyrmion}) it follows that this continuous
$N=1$ vortex can be also represented as the vortex in the
$|\uparrow\rangle$  component whose core is filled by the
$|\downarrow\rangle$  component.

The nematic state (ii) may contain a no less exotic topological object --
the topologically stable $N=1/2$ vortex  \cite{VolMinTop} -- which still has
avoided experimental identification in superfluid $^3$He-A.
The $N=1/2$ vortex is a combination of the $\pi$-vortex in the phase $\Phi$
and $\pi$-disclination in the nematic  order parameter vector $\hat{\bf d}$:
\begin{equation}
{\bf a}=f(r)\left(\hat{\bf x} \cos{\phi\over 2} +\hat{\bf y}
\sin{\phi\over 2}\right)e^{i\phi/2}~.
\label{HalfQuantumVortex}
\end{equation}
The change of the sign of the vector $\hat{\bf d}$ when circumscribing
around the
core is compensated by the change of sign of the exponent
$e^{i\Phi}=e^{i\phi/2}$, so that the whole order parameter is smoothly
connected after circumnavigating.

This $N=1/2$ vortex is the counterpart of the so called Alice string
considered in
particle physics\cite{Schwarz}: a particle which moves around an Alice string
flips its charge or parity.  In a similar manner a quasiparticle adiabatically
moving around the vortex in
$^3$He-A or in the Bose condensate with $F=1$ in nematic state (ii) finds its
spin or its momentum projection $M$ reversed with respect to the fixed
environment.   This is because the
$\hat {\bf d}$-vector, which plays the role of the quantization axis for
the spin of a quasiparticle, rotates by $\pi$ around the vortex.
As a consequence, several phenomena ({\it e.g.} global Aharonov-Bohm effect)
discussed in the particle physics literature \cite{March-Russel1992,Davis1994}
correspond to effects in $^3$He-A  physics
\cite{Khazan1985,SalomaaVolovik1987}, which can be extended to the atomic Bose
condensates.

In high-temperature superconductors with a nontrivial order
parameter, the half-quantum vortex was identified as being attached to the
intersection line of three grain boundaries \cite{Kirtley1996}, as
suggested in \cite{Geshkenbein1987}.  This $N=1/2$ vortex has been observed via
the fractional magnetic flux it generates.

In the spin projection representation the order parameter asymptote in the
$N=1/2$ vortex in the nematic phase is
\begin{equation}
\Psi_\nu\propto e^{i\phi/2}\left(
\matrix{
e^{i\phi/2}\cr
0\cr
e^{-i\phi/2}\cr
}\right)=\left(
\matrix{
e^{i\phi}\cr
0\cr
1\cr
}\right)
\,\,.
\label{VortexInOneComponent}
\end{equation}
This means that the $N=1/2$ vortex can be represented as a vortex in the
spin-up component $|\uparrow\rangle$, while the spin-down component
$|\downarrow\rangle$ is vortex-free. Such representation of the half-quantum
vortex in terms of the regular $N=1$ vortex in one of the components of the
order
parameter occurs also in the $^3$He-A.
The general form of the order parameter in the half-quantum vortex, which
includes also the core structure, is
\begin{equation}
\Psi_\nu= \left(
\matrix{
f_1(r)e^{i\phi}\cr
0\cr
f_2(r)\cr
}\right)~~,~~f_1(0)=0~~, ~~|f_1(\infty)|=|f_2({\infty}) |~~.
\label{VortexInOneComponentGeneral}
\end{equation}

Note that, since the $M=0$ component in Eq.(\ref{VortexInOneComponent}) is
zero,
the half-quantum vortex can be generated also in the Bose-condensate with two
internal degrees of freedom, explored in Ref. \cite{Matthews}. The necessary
condition for that is that in the equilibrium state of such condensate both
components must be equally populated. This is required by the asymptote of
Eq.(\ref{VortexInOneComponentGeneral}), where both components have the same
amplitude. If the amplitudes are not exactly equal, the half-quantum vortex
acquires the tail in the form of the domain wall terminating on the vortex. The
same happens in $^3$He-A where the half-quantum vortex is the termination line
of the topological soliton.

The Eq.(\ref{VortexInOneComponent}) may suggest a way to generate a
half-quantum vortex in an alkali Bose--Einstein condensate,
simply be combining the successful idea \cite{Matthews,Williams}
for producing skyrmions with the proposal \cite{Dobrek}
for making scalar vortices by the effect of light forces.
Let us start from the homogeneous state
\begin{equation}
\Psi_\nu({\rm initial})=f  \left(
\matrix{
e^{i\alpha}\cr
0\cr
e^{i\beta}\cr
}\right)
\label{InitialState}~,
\end{equation}
which corresponds to the phase $\Phi=(\alpha +\beta)/2$ and the nematic
vector $\hat{\bf d}=\hat{\bf x}\cos (\alpha -\beta)/2 +\hat{\bf y}\sin
(\alpha -\beta)/2$.
A light spot shall illuminate the condensate with an intensity
distribution $I$ that draws a half-quantized vortex,
\begin{equation}
I = I_0 e^{i\phi/2}
\,\,.
\end{equation}
The light should be a short pulse and it should be non--resonant with
respect to atomic transition frequencies.
Simultaneously, uniform microwave radiation shall penetrate the condensate.
The radiation should be far--detuned from the transition frequency
between the spin components $|\downarrow\rangle$ and $|\uparrow\rangle$ of the
condensate such that it only causes shifts in the relative phases
between $|\downarrow\rangle$ and $|\uparrow\rangle$ and no population transfer.
The light spot will imprint an optical mask onto the homogeneous
microwave field, due to the optical Stark effect.
Therefore, the generated relative phase shift will follow the half-quantum
vortex drawn by the light spot.
Simultaneously, the condensate gains an overall scalar phase factor,
caused by the intensity kick of the light.
This factor should exactly compensate the phase mismatch between the
components that is left from the optically assisted microwave effect.
Of course, for this the intensities of light and microwave radiation
should be properly adjusted, but this could be arranged.
In this simple way, an Alice string can be created in a
multicomponent Bose--Einstein condensate of alkali atoms.

{\bf Acknowledgements.}

We thank Matti Krusius and Brian Anderson for fruitful discussions. The work of
GEV was supported in part by the
Russian Foundations for Fundamental Research
and by
European Science Foundation.
UL was supported by the Alexander von Humboldt Foundation and the
G\"oran Gustafsson Stiftelse.


\begin{thebibliography}{15}

\bibitem{Monopoles} G.E. Volovik,  Proc. Natl. Acad. Sc. USA {\bf 97},
2431 (2000).

\bibitem{Ho} T.L. Ho,   Phys. Rev. Lett. {\bf 81}, 742 (1998).

\bibitem{TheoryVortexFormation}  T. Isoshima, M. Nakahara, T. Ohmi \& K.
Machida, "Creation of persistent current and vortex in a Bose-Einstein
condensate
of alkali-metal atoms",  cond-mat/9908470; T. Ohmi \& K. Machida
"Bose-Einstein  condensation with internal degrees of freedom in alkali atom
gases" J. Phys. Soc. Jpn. {\bf  67}, 1822 (1998).

\bibitem{Blaauwgeers}  R.~Blaauwgeers, V.~B.~Eltsov, H.~G\"otz,
  M.~Krusius, J.J.~Ruohio, R.~Schanen, and  G.~E.~Volovik, "Double-quantum
vortex in superfluid $^3$He-A",   Nature, March 30 (2000).

\bibitem{EltsovKrusius}   V.~B.~Eltsov \&    M.~Krusius, "Topological
defects in
$^3$He superfluids",  in "Topological Defects and the Non-Equilibrium
Dynamics of Symmetry Breaking Phase Transitions", Eds. Y.M. Bunkov and H.
Godfrin, Kluwer Academic Publishers, 2000, pp. 325-344.

\bibitem{SalomaaVolovik1987} M. M. Salomaa  \&  G.E. Volovik,   Rev. Mod.
Phys.  {\bf 59}, 533 (1987).

\bibitem{Mermin-Ho} N.D. Mermin  \& T.L. Ho,  Circulation and angular
  momentum in the A phase of superfluid $^3$He, Phys. Rev. Lett., {\bf
    36}, 594 (1976).

\bibitem{Marzlin} K.P. Marzlin, W. Zhang \& B.C. Sanders
"Creation of Skyrmions in a Spinor Bose-Einstein Condensate",
cond-mat/0003273.

\bibitem{HoTalk} T.L.  Ho, "Bose-Einstein Condensates with Internal
Degrees of Freedom", invited talk at International Conference on Low
Temperature Physics LT-22 (Helsinki, August 1999).


\bibitem{HindmarshKibble}  Hindmarsh, M. \&  Kibble, T.W.B. Cosmic Strings
   Rep. Prog. Phys. {\bf 58}, 477 (1995).

\bibitem{AchucarroVachaspati}  A. Achucarro  \& T. Vachaspati, "Semilocal and
Electroweak Strings", hep-ph/9904229 to be published in   Phys. Rep. (2000).

\bibitem{Matthews} M. R. Matthews, B. F. Anderson, P. C. Haljan,
D. S. Hall, C. E Wieman, and E. A. Cornell,
Phys. Rev. Lett. {\bf 83}, 2498  (1999).

\bibitem{Williams} J. E. Williams, \& M. J. Holland,  "Preparing Topological
States of a Bose-Einstein Condensate",
Nature {\bf 401}, 568 (1999).

\bibitem{VolMinTop} G.E. Volovik  \&  V.P. Mineev,  JETP Lett. {\bf 24}, 561
(1976).

\bibitem{Schwarz}   A.S. Schwarz,  Nucl. Phys.  {\bf B~208},
141  (1982).

\bibitem{March-Russel1992} J. March-Russel,  J.    Preskill  \&  F. Wilczek,
 Phys. Rev.    {\bf D~ 50}, 2567 (1992).

\bibitem{Davis1994}  A.C. Davis \&  A.P.   Martin,  Nucl. Phys. {\bf B~419},
341 (1994).

\bibitem{Khazan1985}  M.V. Khazan,  JETP Lett. {\bf 41}, 486 (1985).



\bibitem{Kirtley1996} J.R. Kirtley,  C.C. Tsuei, M. Rupp,   et al.,
Phys. Rev. Lett., {\bf 76}, 1336 (1996).

\bibitem{Geshkenbein1987} V. Geshkenbein, A. Larkin  \&  A. Barone,
Phys. Rev.  {\bf B ~36}, 235 (1987).


\bibitem{Dobrek}  L. Dobrek, M. Gajda, M. Lewenstein,
K. Sengstock, G. Birkl \& W. Ertmer,   Phys. Rev. {\bf A~ 60},
R3381  (1999).

\end{thebibliography}
\end{document}